\documentclass[aps,pra,floatfix,twocolumn,superscriptaddress]{revtex4-2}

\usepackage{mathrsfs}
\usepackage{graphicx}
\usepackage[english]{babel}
\usepackage{amsmath}
\usepackage{amssymb}

\usepackage{relsize}
\usepackage{CJK}
\usepackage{dsfont}
\usepackage{bm}

\renewcommand{\parallel}{\mathrel{/\mskip-4mu/}}

\newcommand{\me}{\mathrm{e}}
\newcommand{\mi}{\mathrm{i}}

\allowdisplaybreaks
\begin{document}

\title{Metamorphic dynamical quantum phase transition in double-quench processes at finite temperatures}

\author{Xu-Yang Hou}
\affiliation{School of Physics, Southeast University, Jiulonghu Campus, Nanjing 211189, China}
\author{Qu-Cheng Gao}
\affiliation{School of Physics, Southeast University, Jiulonghu Campus, Nanjing 211189, China}
\author{Hao Guo}
\email{guohao.ph@seu.edu.cn}
\affiliation{School of Physics, Southeast University, Jiulonghu Campus, Nanjing 211189, China}
\author{Chih-Chun Chien}
\email{cchien5@ucmerced.edu}
\affiliation{Department of physics, University of California, Merced, CA 95343, USA}
\begin{abstract}
By deriving a general framework and analyzing concrete examples, we demonstrate a class of dynamical quantum phase transitions (DQPTs) in one-dimensional two-band systems going through double-quench processes. When this type of DQPT occurs, the Loschmidt amplitude vanishes and the rate function remains singular after the second quench, meaning the final state continually has no overlap with the initial state. This type of DQPT is named metamorphic DQPT to differentiate it from ordinary DQPTs that only exhibit zero Loschmidt amplitude and singular rate function at discrete time points. The metamorphic DQPTs occur at zero as well as finite temperatures. Our examples of the Su-Schrieffer-Heeger (SSH) model and Kitaev chain illustrate the conditions and behavior of the metamorphic DQPT. Since ordinary DQPTs have been experimentally realized in many systems, similar setups with double quenches will demonstrate the metamorphic DQPT. Our findings thus provide additional controls of dynamical evolution of quantum systems.
\end{abstract}

\maketitle
\section{Introduction}
With the rapid development in quantum technology, dynamic behaviors of isolated quantum systems have attracted huge research interest~\cite{ZwergerRMP08,PolkovnikovRMP11,Dziarmaga10,EisertNP15,Chien15}. Among many interesting phenomena in nonequilibrium physics, dynamical quantum phase transitions (DQPTs) \cite{Zvyagin16,DQPT13,DQPTreview18} have emerged as a thriving field in atomic, molecular, and optical physics and condensed matter physics. The importance of DQPTs lies in their direct relations to observable behaviors of quantum many-body systems in quench dynamics. Moreover, the studies of DQPTs have shed light on fundamentals of quantum physics and added great controls to our quantum toolbox.

In a common definition, a DQPT reveals nonanalytic behavior in the real-time dynamics of a quantum system. It has been formulated in Ref. \cite{DQPT13} in the context of quantum dynamics following a quench. By an analogy of nonanalytic behavior in thermodynamic phase transition, DQPT provides an elegant description of a class of nonequilibrium phase transitions~\cite{DQPT14,DQPT15,DQPTreview18}. There
have been extensive research in theoretical studies \cite{DQPTB2,DQPTB3,DQPTB4a0,DQPTB4a,DQPTB4b,DQPTB4c,DQPTB4d0,DQPTB4d,DQPTB4d2,DQPTB4e1,DQPTB4e,DQPTB5,JafariSciRep19}, as well as pioneering experiments realizing DQPTs in ionic and atomic
systems \cite{DQPTB41,DQPTB4}.

The analogy of the free energy in thermodynamic phase transitions is the rate function in a DQPT for describing the post-quench behavior of a quantum system. Another important
quantity in the study of DQPTs is the return amplitude, also known as the Loschmidt amplitude, which is a function of time $t$ and acts as the
analog of the partition function in thermodynamics. While a thermodynamic phase transition occurs when the free energy exhibits singular behavior, a DQPT occurs at critical times $t^*_n$ that are zeros of the Loschmidt amplitude~\cite{DQPTreview18} and cause singular behavior in the rate function. The zeros are called the Fisher zeros when $t$ is complexified~\cite{DQPT13}. The appearance of the critical times signals that the pre- and post-quench quantum states become orthogonal and belong to different quantum phases. Furthermore, crossing a quantum critical points by a quantum quench may also induce DQPTs~\cite{DQPTreview18}.

While early studies
of DQPTs focused on pure quantum states and their dynamics, there have been attempts to generalize the concept of DQPT to mixed quantum states~\cite{DQPTM16,DTQPT18,DQPTM17,DQPTViyula18,DQPTLang18,ourPRB20b}. A primary method is to extend the concept of Loschmidt amplitude in a suitable way to finite temperatures using the density matrix~\cite{DQPTreview18,Sedlmayr18}. Since the system after a quench is usually not in equilibrium, the temperature is referred to that of the initial state in equilibrium.
Moreover, there have been studies of DQPTs in topological systems~\cite{TDQPT1,TDQPT2,TDQPT3,TDQPT4,TDQPT5,TDQPT6} after a single quench and quench dynamics in other context~\cite{Jafari19,Mishra20,Sadrzadeh21}.

More recently, DQPTs in multiple-quench processes have been proposed~\cite{DQPRB18} to go beyond the single-quench processes in most of the studies of DQPTs.
It has been shown that in a double-quench process, both the absence
or presence of nonanalytic behavior before and after the second
quench can be demonstrated in a quantum system, thereby providing additional controls of DQPTs.
Here, we generalize the formalism of DQPTs in double-quench processes to finite temperatures since mixed quantum states are common in the real world. To generalize the physical quantities involved in DQPTs to mixed states, we adopt the idea of defining the Loschmidt amplitude as the overlap between the purified states of the density matrices~\cite{DTQPT18}. We will show that by tuning the time duration between the two quenches, the quantum system can be driven to a final state that is maximally different from (orthogonal to) the initial state and will never regain its overlap with the initial state after the last DQPT at the second quench. We refer to this type of DQPT as the metamorphic DQPT to distinguish it from the ordinary DQPT from a single quench that only shows discrete singular points. Since the framework is general, the metamorphic DQPT can occur in pure quantum states going through double-quench processes as well.

The rest of the paper is organized as follows. In Sec. \ref{2}, we give a general introduction to the setup and framework for investigating DQPTs in double-quench processes. A detailed theoretical analysis then shows where ordinary and metamorphic DQPTs will emerge. Section \ref{3} presents two explicit examples, the Su-Schrieffer-Heeger (SSH) model and the Kitaev chain, to demonstrate the metamorphic DQPTs and discuss the behavior associated with various DQPTs. Sec.~\ref{sec:4} discusses some implications for experiments. Finally, Sec.~\ref{sec:con} concludes our work.

\section{Theoretical Framework and method}\label{2}
We first give a briefly overview of DQPTs. In the following, we will set $\hbar=1=k_B$. The key object in the theory of DQPTs is the Loschmidt (or return) amplitude~\cite{DQPT13,DQPTreview18}
\begin{align}\label{LA}
\mathcal{G}(t)=\langle\psi(0)|\psi(t)\rangle=\langle\psi(0)|\me^{-\mi Ht}|\psi(0)\rangle,
\end{align}
where $t$ represents time, $|\psi(0)\rangle$ is the initial quantum state, and $H$ denotes the quenched Hamiltonian. The Loschmidt amplitude measures the deviation of the time-evolved state from the initial condition, and its zeros $t^*_n$ denote the transition points of DQPTs~\cite{DQPTM17,DTQPT18}. While a quantum phase transition~\cite{Sachdev_book} occurs when the ground states become orthogonal across a critical point determined by a parameter in the Hamiltonian, a DQPT similarly describes orthogonality of states induced by time evolution. Since a dynamical quench process is generically not an equilibrium process, rather than the conventional thermodynamic free-energy, the ``dynamical version'' of the free-energy is introduced as
 \begin{align}\label{RF}
f(t)=-\frac{1}{N}\lim_{N\rightarrow \infty}\ln|\mathcal{G}(t)|^2,
\end{align}
where $N$ is the overall degrees of freedom.
Accordingly, $|\mathcal{G}(t)|^2$ plays the role of the partition function in thermodynamics~\cite{DQPTreview18}.
$f(t)$ is usually called the rate function, which exhibits singular behavior at the zeros of $\mathcal{G}(t)$.

In quantum information theory~\cite{WatrousBook}, $\mathcal{G}(t)$ is referred to as the quantum fidelity, qualifying the ``similarity'' between the initial and final states. Uhlmann defined two pure states to be parallel if and only if the fidelity between them is a positive real number~\cite{Uhlmann86}: $\langle \psi_1|\psi_2\rangle=\langle \psi_2|\psi_1\rangle>0$. Using this definition, if two states are perpendicular to each other with $\langle \psi_1|\psi_2\rangle=0$, they are said to have minimal similarity since the state vectors $|\psi_{1,2}\rangle$ contain no component of each other. Therefore, the evolved quantum state will have minimal similarity with respect to the initial state at $t^*_n$s, leading to DQPTs.

To generalize the formalism of DQPTs to mixed quantum states, it is convenient if we have a pure-state like description of mixed states. This can be achieved by a protocol called purification of density matrices~\cite{Uhlmann86}. Suppose a mixed state is described by the density matrix $\rho=\sum_i\lambda_i|i\rangle\langle i|$. If $\rho$ has full rank, a matrix $W=\sqrt{\rho}U=\sum_i\sqrt{\lambda_i}|i\rangle\langle i|U$ with $U$ being an arbitrary unitary matrix is said to purify $\rho$ since $\rho=WW^\dagger$. $W$ is called the purification or amplitude of $\rho$. The purified state given by
$|W\rangle=\sum_i\sqrt{\lambda_i}|i\rangle\otimes U^T|i\rangle$ is isomorphic to $W$. The inner product of two purified states follows the Hilbert-Schmidt product $\langle W_1|W_2\rangle=\text{Tr}(W^\dagger_1 W_2)$.

The concept of the Loschmidt amplitude can be generalized as follows. If the initial mixed state of a quantum system is given by $\rho(0)=W(0)W^\dag(0)$, the density matrix will evolve as $\rho(t)=\me^{-\mi Ht}\rho(0)\me^{\mi Ht}$ after a quench governed by $H$. This leads to $W(t)=\me^{-\mi Ht}W(0)$. Thus, the Loschmidt amplitude can be obtained by generalizing Eq. (\ref{LA}) to
 \begin{align}
 \mathcal{G}_\rho(t)
 &=\langle W(0)|W(t)\rangle=\text{Tr}\left[W^\dag(0)W(t)\right]\notag\\
 &=\text{Tr}\left[\rho(0)\me^{-\mi Ht}\right].
\end{align}
We mention that equivalent expressions have been found as the transition amplitude of purified states~\cite{DTQPT18,ourPRB20b}. Similarly, the generalization of the rate function (\ref{RF}) is given by
 \begin{align}\label{gt2}
 g(t)=-\lim_{N\rightarrow \infty}\frac{1}{N}\log | \mathcal{G}_\rho(t)|^2.
\end{align}
Here the limit $N\rightarrow\infty$ must be taken at the end of the evaluation. As pointed out previously~\cite{DQPTreview18}, at critical times $t^*_n$, the rate function shows nonanalytic
behavior if the quench process
induces a dynamical quantum critical point, where the Loschmidt amplitude vanishes.

To give concrete examples of dynamical effects of quench processes at finite temperatures, we consider a generic one-dimensional (1D) two-band Hamiltonian with periodic boundary condition $H=\sum_k\Psi^\dagger_k H_k\Psi_k $ relevant to a group of systems in condensed matter physics. Here $\Psi_k$ is a Nambu spinor of a pair of fermionic operators, and $H_k$ is a $2\times 2$ matrix of the form
\begin{align}\label{Hk}
H_{k}=E_{k}+\frac{1}{2}\Delta_{k}\hat{n}_{k}\cdot\vec{\sigma},
\end{align}
where $\vec{\sigma}=(\sigma_x,\sigma_y,\sigma_z)$ are the Pauli matrices, $\Delta_k$ corresponds to the energy gap of $H_k$, and the function $E_k$ plays no important role in the following discussion. The 1D momentum $k$ is limited to the first Brillouin zone and is thus periodic, and the unit vector
$\hat{n}_{k}=(\sin\theta_k\cos\phi_k,\sin\theta_k\sin\phi_k,\cos\theta_k)^{T}$. The reason we choose this type of models is because allow exact solutions, many of which exhibit interesting topological properties, as we will see shortly. Moreover, those 1D two-band models have minimal numbers of controllable parameters for the metamorphic DQPT in double-quench processes, allowing succinct demonstrations.

While a typical DQPT involves only one quench, here we consider a double-quench process governed by $H(t)$. At $t=0$, the system experiences the first quench, and the Hamiltonian is suddenly switched from $H_0$ to $H_{1}$. At a later time $t=\tau$, a second quench is applied, and the Hamiltonian is subsequently switched to $H_2$.
Similar to Eq. (\ref{Hk}), the time-dependent Hamiltonian can be expressed as
\begin{equation}\label{H012}
H_k(t)=
\left\{
  \begin{array}{ll}
H_{0k}=E_{0k}+\frac{1}{2}\Delta_{0k}\hat{n}_{0k}\cdot\vec{\sigma}, &  t<0,\\
H_{1k}=E_{1k}+\frac{1}{2}\Delta_{1k}\hat{n}_{1k}\cdot\vec{\sigma}, &  0\leqslant t <\tau,\\
H_{2k}=E_{2k}+\frac{1}{2}\Delta_{2k}\hat{n}_{2k}\cdot\vec{\sigma}, &   t \geqslant\tau,\\
  \end{array}
  \right.
\end{equation}
where $\tau$ is the time interval between the two quenches.
If $H_{1k}=H_{2k}$ (i.e. $\hat{n}_{1k}=\hat{n}_{2k}$), the model actually reduces to a single-quench process, which has already been extensively studied \cite{DQPTM17,DTQPT18,ourPRB20b}.

To introduce the concept of temperature in quench processes, the initial mixed state $\rho(0)$ is chosen as the thermal equilibrium state at temperature $T$. Using Eq. (\ref{H012}), the corresponding density matrix is
\begin{align}\label{r0}
\rho(0)&=\prod_k\otimes\rho_k(0)=\prod_k\otimes\frac{\me^{-\beta H_{0k}}}{\text{Tr}(\me^{-\beta H_{0k}})} \notag \\
&=\prod_k\otimes\frac{1}{2}\left(1-\tanh\frac{\beta\Delta_{0k}}{2}\vec{\sigma}\cdot\hat{n}_{0k}\right),
\end{align}
where $\beta=\frac{1}{k_B T}$.

After constructing the initial and final purified states, the Loschmidt amplitude as a function of time for the double-quench process is obtained as
\begin{align}\label{los}
    \mathcal{G}_\rho(t)
    =
    \left\{
  \begin{array}{lcl}
    \text{Tr}\left(\rho(0)\me^{-\frac{\mi}{\hbar}H_1 t}\right), &  & 0\leqslant t <\tau,\\
    \text{Tr}\left(\rho(0)\me^{-\frac{\mi}{\hbar}H_2(t-\tau)}\me^{-\frac{\mi}{\hbar}H_1\tau}\right), &  & t \geqslant\tau,\\
  \end{array}
  \right.
\end{align}
where
\begin{align}\label{H12}
    \me^{-\frac{\mi}{\hbar}H_1\tau}&=\prod_k\otimes[\me^{-\frac{\mi}{\hbar}E_{1k}\tau}    (\cos(\omega_{1k} \tau) 1_{2\times 2} \notag \\
   &-\mi\sin(\omega_{1k} \tau)\hat{n}_{1k}\cdot\vec{\sigma})],\notag\\
  \me^{-\frac{\mi}{\hbar}H_2(t-\tau)}&=\prod_k\otimes\me^{-\frac{\mi}{\hbar}E_{2k}(t-\tau)}    (\cos\omega_{2k}(t-\tau) 1_{2\times 2} \notag \\
  &-\mi\sin\omega_{2k}(t-\tau)\hat{n}_{2k}\cdot\vec{\sigma})]
\end{align}
with $\omega_{1,2k}=\frac{\Delta_{1,2k}}{2\hbar}$.
To simplify the notations, we introduce $a_{1k}=\cos(\omega_{1k} \tau)$, $a_{2k}=\cos\omega_{2k} (t-\tau)$, $b_{1k}=\sin(\omega_{1k} \tau)$, and $b_{2k}=\sin\omega_{2k} (t-\tau)$. Plugging Eqs. (\ref{r0}) and (\ref{H12}) into Eq. (\ref{los}) and ignoring any terms linear with respect to the Pauli matrices since $\text{Tr}\sigma_x=\text{Tr}\sigma_y=\text{Tr}\sigma_z=0$, we get
\begin{align}\label{los1a}
    \mathcal{G}_\rho(t)&=\prod_k\frac{C'_k}{2}\left[\cos(\omega_{1k}t)-\mi\sin(\omega_{1k}t)\vec{n}_{k}\cdot\hat{n}_{1k}\right],
\end{align}
for $ 0\leqslant t <\tau$, and
\begin{align}\label{los1}
    \mathcal{G}_\rho(t)&=
     \prod_k\frac{C_k}{2}[a_{1k}a_{2k}-b_{1k}b_{2k}\hat{n}_{1k}\cdot\hat{n}_{2k}-\mi a_{1k}b_{2k}\vec{n}_{k}\cdot\hat{n}_{2k} \notag \\
     &-\mi a_{2k}b_{1k}\vec{n}_{k}\cdot\hat{n}_{1k}+b_{1k}b_{2k}\vec{n}_k\cdot(\hat{n}_{1k}\times\hat{n}_{2k})]
\end{align}
for $t \geqslant\tau$. Here $C'_k=\me^{-\frac{E_{1k}}{\hbar}t}$, $C_k=\me^{-\frac{E_{1k}}{\hbar}\tau}\me^{-\frac{E_{2k}}{\hbar}(t-\tau)}$, and $\vec{n}_k=-\tanh\frac{\beta\Delta_{0k}}{2}\hat{n}_{0k}$ is temperature-dependent.

When $ 0\leqslant t <\tau$, only the first quantum quench is implemented and the result is exactly that of the single-quench process studied previously~\cite{DTQPT18,ourPRB20b}.
According to Eq. (\ref{los1a}), DQPTs may occur at
\begin{align}\label{t1n}
t^*_n=\frac{1}{\omega_{1k_\text{c}}}\left(n\pi+\frac{\pi}{2}\right),
\end{align}
where $n$ is a nonnegative integer (similarly hereinafter) and $k_\text{c}$ is the critical momentum such that $\vec{n}_{k_\text{c}}\cdot\hat{n}_{1k_\text{c}}=0$.

For a genuine double-quench process, one usually has $\hat{n}_{1k}\neq\hat{n}_{2k}$. Inspired by previous discussions for $0\leqslant t<\tau$, we consider the case that there exist another critical momentum $\tilde{k}_\text{c}$ such that $\hat{n}_{1\tilde{k}_\text{c}}\cdot\hat{n}_{2\tilde{k}_\text{c}}=0$. Thus, Eq. (\ref{los1}) implies
\begin{align}\label{losft}
    \mathcal{G}_\rho(t)
    &=\prod_{k\neq \tilde{k}_{\text{c}}}\mathcal{G}^{k}_\rho(t)\times\frac{C_{\tilde{k}_\text{c}}}{2}\Big[\cos(\omega_{1\tilde{k}_\text{c}} \tau)\cos\omega_{2\tilde{k}_\text{c}} (t-\tau) \notag \\
    &-\mi \cos(\omega_{1\tilde{k}_\text{c}} \tau)\sin\omega_{2\tilde{k}_\text{c}} (t-\tau)\vec{n}_{\tilde{k}_\text{c}}\cdot\hat{n}_{2\tilde{k}_\text{c}}  \notag \\
    &-\mi \cos\omega_{2\tilde{k}_\text{c}} (t-\tau)\sin(\omega_{1\tilde{k}_\text{c}} \tau)\vec{n}_{\tilde{k}_\text{c}}\cdot\hat{n}_{1\tilde{k}_\text{c}} \notag \\
    &+\sin(\omega_{1\tilde{k}_\text{c}} \tau) \sin\omega_{2\tilde{k}_\text{c}} (t-\tau)\vec{n}_{\tilde{k}_\text{c}}\cdot(\hat{n}_{1\tilde{k}_\text{c}}\times\hat{n}_{2\tilde{k}_\text{c}})\Big].
\end{align}
Moreover, if we choose a special initial state satisfying $\vec{n}_{\tilde{k}_\text{c}}\parallel\hat{n}_{2\tilde{k}_\text{c}}$ at $\tilde{k}_\text{c}$, then $\vec{n}_{\tilde{k}_\text{c}}\cdot\hat{n}_{1\tilde{k}_\text{c}}=0$ and $\vec{n}_{\tilde{k}_\text{c}}\cdot(\hat{n}_{1\tilde{k}_\text{c}}\times\hat{n}_{2\tilde{k}_\text{c}})=0$, and Eq. (\ref{losft}) can be simplified as
\begin{align}\label{GT0}
    \mathcal{G}_\rho(t)
    &=\frac{C_{\tilde{k}_\text{c}}}{2}\cos(\omega_{1\tilde{k}_\text{c}} \tau)[\cos\omega_{2\tilde{k}_\text{c}} (t-\tau) \notag \\
    &-\mi \sin\omega_{2\tilde{k}_\text{c}} (t-\tau)\vec{n}_{\tilde{k}_\text{c}}\cdot\hat{n}_{2\tilde{k}_\text{c}}]\prod_{k\neq \tilde{k}_{\text{c}}}\mathcal{G}^{k}_\rho(t).
\end{align}
Note the condition $\vec{n}_{\tilde{k}_\text{c}}\cdot\hat{n}_{1\tilde{k}_\text{c}}=0$ actually implies $\tilde{k}_\text{c}=k_\text{c}$, hence we will omit the `hat' of $\tilde{k}_\text{c}$ hereafter. According to Eq. (\ref{GT0}), an interesting result is that $\mathcal{G}_\rho(t)=0$ for any time $t>\tau$ if the interval $\tau$ between the two quenches satisfies \begin{align}\tau=\tau^*=\frac{1}{\omega_{1k_\text{c}} }\left(n\pi+\frac{\pi}{2}\right).\end{align}

The behavior of the rate function when $t>\tau^*$ needs a detailed analysis. For simplicity, we suppose there is only one critical momentum $k_\text{c}$. Using Eq. (\ref{gt2}), we get
\begin{align}\label{gt}
g(t)&
=-\lim _{N \rightarrow \infty} \frac{2}{N} \Big[\ln\frac{C_{k_\text{c}}}{2}+\ln|\cos(\omega_{1k_\text{c}} \tau^*)| \notag \\
+\ln&\left(\cos^2\omega_{2k_\text{c}} (t-\tau^*)+\sin^2\omega_{2k_\text{c}} (t-\tau^*)(\vec{n}_{k_\text{c}}\cdot\hat{n}_{2k_\text{c}})^2\right)\Big]\notag\\
&-\lim _{N \rightarrow \infty} \frac{2}{N} \sum_{k\neq k_\text{c }}\ln|\mathcal{G}^{k}_\rho(t)|.
\end{align}
Importantly, there is always a $t$-independent singular term $\ln\cos(\omega_{1k_\text{c}} \tau^*)$ if the time duration $\tau$ between the two quenches is chosen properly. To understand this, we first explain the meaning of $\mathcal{G}_\rho(t>\tau^*)=0$.
Physically,
$\mathcal{G}_\rho(t^*)=0$ of a single-quench process means that the initial and final mixed states share minimal similarity at $t^*$, as pointed out previously~\cite{ourPRB20b}. For a double-quench process exhibiting $\mathcal{G}_\rho(t>\tau^*)=0$, the condition $\hat{n}_{1k_\text{c}}\cdot\hat{n}_{2k_\text{c}}=0$ can be understood as an implication that $H_1$ and $H_2$ are perpendicular to each other at least at one critical momentum $k_\text{c}$, denoted by $H_1\bot H_2$ at $k_\text{c}$ hereafter. Thus, a DQPT occurs when the second quench is applied at $t=\tau^*$ if $H_1\bot H_2$ at $k_\text{c}$. After that, the subsequent dynamical evolution is governed by $H_2$, and the system stays in the state maximally different from the initial state since we always have $ \mathcal{G}_\rho(t)=0$ for $t>\tau^*$ in this case.

We emphasize that this phenomenon cannot happen in the 1D two-band system going through a single-quench process at finite temperatures, which only allows $\mathcal{G}_\rho=0$ at discrete points. Since the final state remains orthogonal to the initial state after the second quench in a double-quench process with suitable parameters, we refer to the DQPT occurring at $t=\tau^*$ as a metamorphic DQPT. Importantly, the rate function remains singular after a metamorphic DQPT occurs in a double-quench process because the final state does not generate any overlap with the initial state after $\tau^*$.

If the system is initially prepared in an equilibrium state at temperature $T$, then
$\vec{n}_k=-\tanh\frac{\beta\Delta_{0k}}{2}\hat{n}_{0k}$. The condition for the occurrence of a metamorphic DQPT only requires $\hat{n}_{0k_\text{c}}\parallel\hat{n}_{2k_\text{c}}$ and $\hat{n}_{2k_\text{c}}\perp\hat{n}_{1k_\text{c}}$ at $k_\text{c}$. Thus, a simple protocol to ensure the occurrence of a metamorphic DQPT is to let the Hamiltonian after the second quench return to the initial Hamiltonian, i.e., $H_{2}=H_0$ (implying $\hat{n}_{0k}=\hat{n}_{2k}$). In this setting, a metamorphic DQPT can occur if at least one critical momentum $k_\text{c}$ exists. In our examples, we will focus on this simple setting.

We emphasize that the metamorphic DQPTs in double-quench processes may also occur if the initial states are pure quantum states.
By substituting $\vec{n}_k=-\tanh\frac{\beta\Delta_{0k}}{2}\hat{n}_{0k}$ into Eq. (\ref{GT0}), we have
\begin{align}\label{losft2}
    \mathcal{G}_\rho(t)&=\frac{C_{k_\text{c}}}{2}\cos(\omega_{1k_\text{c}} \tau)[\cos\omega_{2k_\text{c}} (t-\tau) \notag \\
    &+\mi\sin\omega_{2k_\text{c}} (t-\tau)\tanh\frac{\beta\Delta_{0k_\text{c}}}{2}]\prod_{k\neq k_\text{c}}G^k_\rho(t).
\end{align}
At zero temperature ($\beta\rightarrow\infty$), we have $\lim_{\beta\rightarrow\infty}\tanh\frac{\beta\Delta_{0k}}{2}=1$ and $\rho(0)=\frac{1}{2}(1-\vec{\sigma}\cdot\hat{n}_{0k})$. The latter is just the projection operator of the ground-state energy level $E_{0k}-\frac{1}{2}\Delta_{0k}$. The Loschmidt amplitude reduces to
\begin{align}\label{losft3}
    \mathcal{G}_\rho(t)&=\frac{C_{k_\text{c}}\me^{\mi\omega_{k_{\text{c}}}(t-\tau)}}{2}\cos(\omega_{1k_\text{c}} \tau)\prod_{k\neq k_\text{c}}G^k_\rho(t).
\end{align}
Therefore, metamorphic DQPTs may occur if $\tau=\frac{1}{\omega_{1k_\text{c}}}\left(n\pi+\frac{\pi}{2}\right)$.

Furthermore, we point out that there are no subsequent ordinary DQPTs after the second quench ($t>\tau^*$) is applied in the situation with $\hat{n}_{0k_\text{c}}\parallel\hat{n}_{2k_\text{c}}$ and $\hat{n}_{2k_\text{c}}\perp\hat{n}_{1k_\text{c}}$  at finite temperatures. The only nonanalyticity of $g(t>\tau^*)$ comes from the metamorphic DQPT at $t=\tau^*$, which can be deduced from Eqs. (\ref{losft}) and (\ref{GT0}). We will also verify this observation numerically in our examples. Moreover, for the 1D two-band model analyzed here (with at least one $k_\text{c}$ such that $\hat{n}_{0k_\text{c}}\parallel\hat{n}_{2k_\text{c}}$ and $\hat{n}_{2k_\text{c}}\perp\hat{n}_{1k_\text{c}}$), it can be shown that there is no ordinary DQPT after the second quench if a metamorphic DQPT is absent, i.e., $\tau\neq\tau^*$. Nevertheless, ordinary DQPTs may still appear at $t>\tau$ in more general situations of double-quench processes by controlling $\tau$ and the parameters of $H_0$, $H_1$ and $H_2$, as discussed in Ref. \cite{DQPRB18}.

Our discussion so far focuses on the simple 1D two-band quantum systems.
Here we briefly analyze whether the metamorphic DQPTs are generic phenomena in double-quench processes for other systems. Note that there are at least three controllable parameters ($T$, $\tau$ and $k$) in our analysis based on Eq. (\ref{losft}). The vanishing of $\mathcal{G}_\rho$ only imposes two conditions for the real and imaginary parts. Hence, there are infinite sets of parameters that can fulfill $\mathcal{G}_\rho(t)=0$. At every finite temperature $T$, there are values of $\tau^*$ and $k_\text{c}$ for the metamorphic DQPT. For quantum systems with more than two bands or higher dimensions of the Brillouin zone (with a replacement of $k$ by vector $\mathbf{k}$), there are more controllable parameters and Eq. (\ref{losft}) becomes more complicated. However, it still imposes only two conditions. Thus, there will be more choices to ensure the occurrence of metamorphic DQPTs with $\mathcal{G}_\rho(t)=0$, and we expect more complicated phase diagrams. Nevertheless, the 1D two-band models are valuable because they provide minimal setups with concrete and exact results of the metamorphic DQPTs.

\section{Examples}\label{3}
Here we present two explicit examples to demonstrate the metamorphic DQPTs at finite temperatures in double-quench processes. Instead of repeating the construction of the purified states and the evaluation of the Loschmidt amplitude for locating various DQPTs, here we will use the conditions for generic 1D two-band models and analyze the parameters of the examples that meet the conditions.

\subsection{SSH Model}
The first example is the Su-Schrieffer-Heeger (SSH) model~\cite{SSH}, which is a paradigm of 1D topological insulators. We will see that the metamorphic DQPTs are closely linked to the topological properties of the SSH model.
The SSH model is described by the Hamiltonian with periodic boundary condition:
\begin{equation}
\hat{H}=\sum^{L}_{i=1}(J_{1} a^{\dag}_{i}b_{i}+J_{2} a^{\dag}_{i}b_{i-1}+\text{H.c.} ),
\end{equation}
where the alternating hopping coefficients $J_{1,2}$ are both positive.
The Hamiltonian can be cast into the form (\ref{Hk}) in momentum space with
\begin{align}E_k&=0,\notag\\ \Delta_{k}&=2\sqrt{J^2_{1}+J^{2}_{2}+2J_{1}J_{2}\cos k},\notag\\ \hat{n}_{k}&=\frac{2}{\Delta_{k}}(-J_{1}-J_{2}\cos k,J_{2}\sin k,0)^{T}.\end{align}
In the SSH model, it is actually the dimensionless parameter $\frac{J_1}{J_2}$ that determines the properties of the system.
For a double-quench process, there are three Hamiltonians of the SSH model at different times. The corresponding parameters are respectively labeled by $J_{i1}$ and $J_{i2}$, corresponding to $H_{i}$ ($i=0, 1,2$).

\begin{figure}[th]
\centering
\includegraphics[width=3.0in,clip]{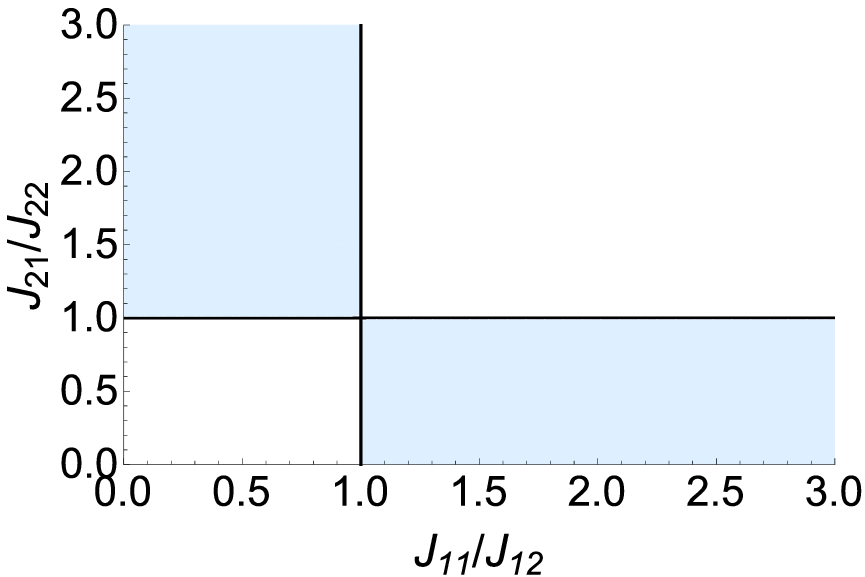}
\caption{Phase diagram of the SSH model in the double-quench process described in the context with the shaded regions indicating where metamorphic DQPTs may be found.}
\label{Fig0}
\end{figure}

A metamorphic DQPT can occur if the critical momentum $k_{\text{c}}$, defined by $\hat{n}_{1k_\text{c}}\cdot\hat{n}_{2k_\text{c}}=0$, can be found. It can be shown that
\begin{align}
\hat{n}_{1k}\cdot\hat{n}_{2k}
&=\frac{4}{\Delta_{1k}\Delta_{2k}}\bigg[(J_{11}+J_{12}\cos k)(J_{21}+J_{22}\cos k) \notag \\
&+J_{12}J_{22}\sin^2 k\bigg].
\end{align}
Let $x=\cos k_\text{c}$, the existence of $k_\text{c}$ requires
\begin{align}\label{x}
J_{11}J_{21}+J_{12}J_{22}+(J_{11}J_{22}+J_{12}J_{21})x=0,
\end{align}
whose root is \begin{align}\label{rox}x=-\frac{J_{11}J_{21}+J_{12}J_{22}}{J_{11}J_{22}+J_{12}J_{21}}.\end{align} The constraint $-1\leqslant x\leqslant1$ needs to be considered.
Note that $x$ is always negative, implying $k_\text{c}>\frac{\pi}{2}$. Using Eq. (\ref{rox}), the condition $x\geqslant -1$ is equivalently expressed as
\begin{align}
(J_{11}-J_{12})(J_{21}-J_{22})\leqslant 0.
\end{align}
In other words, if $J_{11}\leqslant (\geqslant)J_{12}$, then $J_{21}\geqslant (\leqslant)J_{22}$ is required for the existence of $k_c$. The bulk bands of the SSH model with $J_{i1}<J_{i2}$ exhibits different topology from that with $J_{i1}>J_{i2}$~\cite{Asboth2016}, which is related to the Zak phase that can be measured in cold atoms~\cite{Zakphase}. Thus, the condition of $H_1\bot H_2$ at $k_\text{c}$ is equivalent to that of the bulk bands of $H_{1,2}$ having different topologies. This provide a possible way to realize metamorphic DQPTs in experiments: One can manipulate the parameters of the Hamiltonian (the ratio $\frac{J_{i1}}{J_{i2}}$) such that the topological property of the system is changed after each quench. The condition for the existence of the critical momentum in this case is indicated by the shaded regions in Fig. \ref{Fig0} as a phase diagram in terms of the two ratios $\frac{J_{11}}{J_{12}}$ and $\frac{J_{21}}{J_{22}}$.

The last condition to ensure the occurrence of metamorphic DQPTs is that the initial state satisfies $\hat{n}_{0k_\text{c}}\parallel\hat{n}_{2k_\text{c}}$, as discussed before. This further requires $J_{01}J_{22}=J_{02}J_{21}$.
Moreover, the duration between the two quenches that induces a metamorphic DQPT is determined via $\omega_{1k_{\text{c}}}\tau^*=n\pi+\frac{\pi}{2}$. Using Eq. (\ref{x}), we get \begin{align}\label{tau1}\tau^*=\frac{n\pi+\frac{\pi}{2}}{\sqrt{J^2_{1}+J^{2}_{2}+2J_{1}J_{2}\cos k_\text{c}}} .\end{align}
Therefore, after the second quench is applied at $t=\tau^*$ and all those conditions are satisfied, a metamorphic DQPTs always occurs at any finite temperatures.

\begin{figure}[th]
\centering
\includegraphics[width=3.2in,clip]{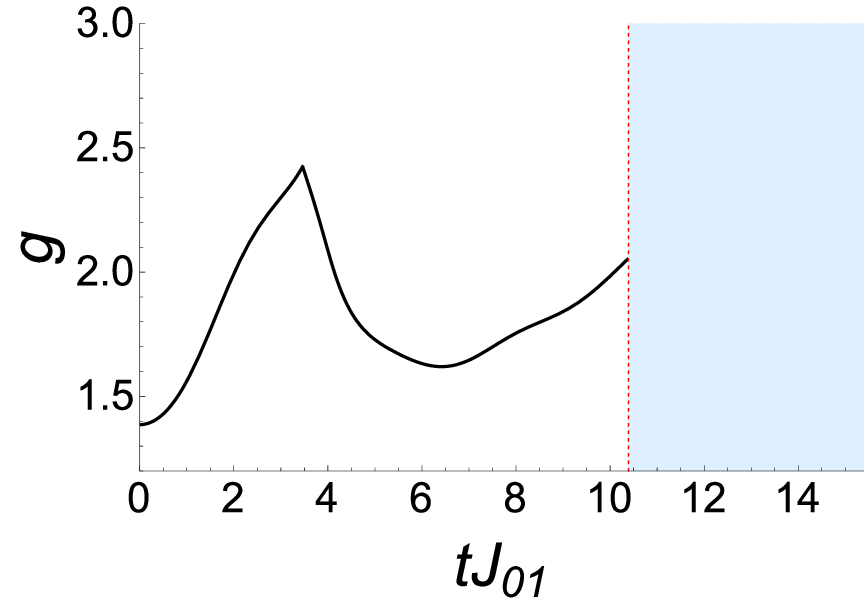}
\includegraphics[width=3.2in,clip]{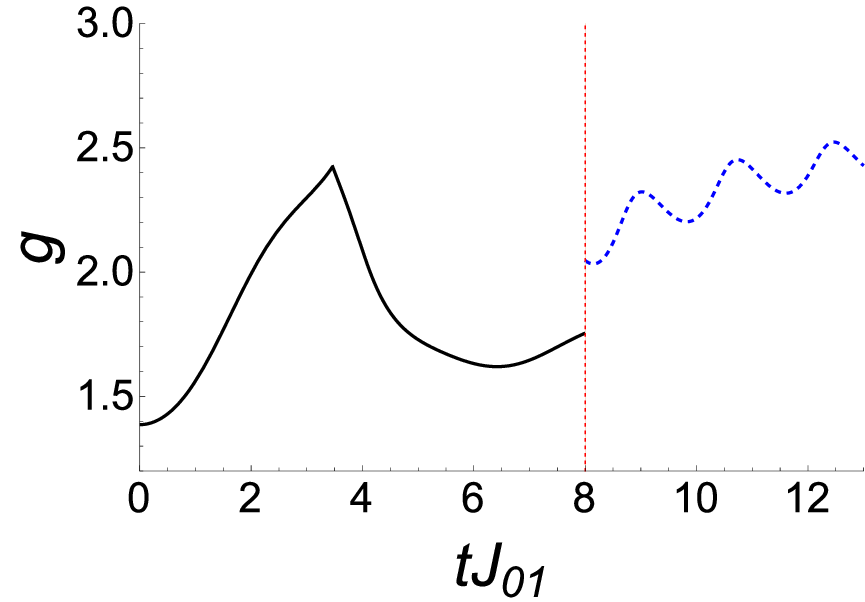}
\caption{Rate function $g(t)$ vs. $t$ for the SSH model going through a double-quench process with $J_{11}=0.4$, $J_{12}=0.8$, $J_{01}=J_{21}=1.0$, and $J_{02}=J_{22}=0.8$ at $T=3.0J_{01}$. (Top panel) $\tau=\tau^*=11.044/J_{01}$ (indicated by the vertical line) with a metamorphic DQPT at $t=\tau^*$. The rate function diverges in the shaded region. (Bottom panel) $\tau=8.0/J_{01}$ (indicated by the vertical line) without a metamorphic DQPT. For the bottom panel, no singular behavior arises for $t>\tau$. The kinks on both panels in the range $0<t<\tau$ indicate the ordinary DQPTs.}
\label{Fig1}
\end{figure}

To visualize our results, we consider an explicit example of the SSH model with $J_{11}=0.4$, $J_{12}=0.8$, $J_{01}=J_{21}=1.0$, and $J_{02}=J_{22}=0.8$ at $T=3.0J_{01}$, and present our numerical results by plotting the rate function $g$ as a function of $t$ in Figure \ref{Fig1}. In the top panel, the duration between the two quenches is chosen as $\tau=\tau^*=11.044/J_{01}$ by setting $n=1$ in Eq. (\ref{tau1}). Hence, when $t<\tau^*$, a DQPT occurs at $t^*_0=3.681/J_{01}$ according to Eq. (\ref{t1n}), which is reflected by the sharp peak of $g(t)$. 
Note that the rate function includes the contributions from $k\neq k_\text{c}$ as well, so the singular behavior is a kink at $t^*_0$ in $g(t)$. At $t=\tau^*$, the system experiences the second quench, which induces a metamorphic DQPT. As a consequence, the rate function is always singular when $t>\tau^*$ in this case, and we use a shaded area to cover that region following the metamorphic DQPT. After the second quench, the dynamical behavior of the rate function except the divergent term $\ln\cos\omega_{1k\text{c}}\tau^*$ is completely analytic in this situation, as we have discussed before. However, the details are concealed by the divergence of $g(t)$ after $t=\tau^*$.

As a comparison, we choose a different value of the time duration with $\tau=8.0/J_{01}$ in the bottom panel, where no metamorphic DQPT occurs when the second quench is applied. Thus, the behavior of the rate function is now visible after the second quench. Usually, there is a discontinuity at $t=\tau$ due to the quench. We remark that all peaks in the region where $t>\tau$ are smooth and no DQPT appears in the latter case. Moreover, the metamorphic DQPT of the SSH model can arise if the double quench process crosses the topological transition point $J_1/J_2=1$ twice as $H_0\rightarrow H_1\rightarrow H_2$. However, the direction of crossing the topological transition point does not matter.

\subsection{Kitaev Chain}
The previous example of the SSH model shows a broad range of parameters for observing the metamorphic DQPT after double quenches. Next, we study a double-quench process of the periodic Kitaev chain modeling 1D p-wave superconductors~\cite{Kitaev}. We remark that DQPTs in fermionic BCS superfluids going through single-quench processes have been studied in Ref.~\cite{Rylands21}. The Hamiltonian of the 1D Kitaev chain is given by
\begin{equation}
\hat{H}=\sum^{L}_{i=1}(-J a^{\dag}_{i}a_{i+1}+M a_{i}a_{i+1}-\frac{\mu}{2}a^{\dag}_{i}a_{i}+\text{H.c.} ),
\end{equation}
where $L$ is the number of sites, $J$ is the hopping coefficient, $\mu$ is the chemical potential, and $M>0$ is the superconducting gap.
We introduce the dimensionless parameters $m=\frac{\mu}{2M}$ and $c=\frac{J}{M}$ that control the model. A comparison with the SSH model shows that there is one more controllable parameter in the Kitaev chain, so we expect more complicated results to follow. We further introduce the Nambu spinor $\Psi_k=(a_k,a^\dagger_{-k})^T$ to write the Hamiltonian with periodic boundary condition in the form (\ref{Hk}) in momentum space with
\begin{align}
E_k&=0,\notag\\
\Delta_{k}&=2M\sqrt{(c\cos k-m)^2+\sin^2 k},\notag\\
\hat{n}_{k}&=\frac{2M}{\Delta_{k}}(0,-\sin k,-m+c\cos k)^{T}.
\end{align}

\begin{figure}[th]
\centering
\includegraphics[width=3.0in,clip]{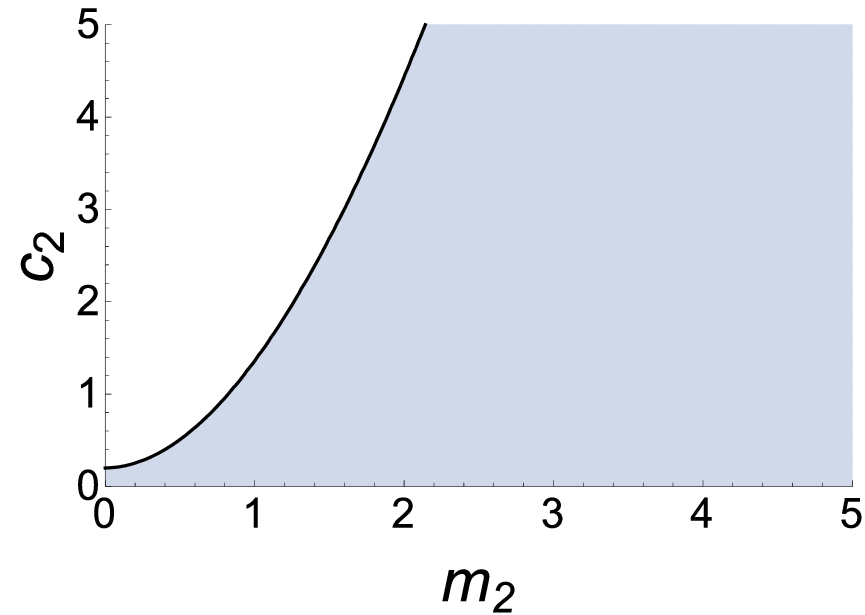}
\caption{$c_2-m_2$ phase diagram with the shaded region indicating where critical momentum exists for the Kitaev chain in a double-quench process with $m_1=0.2$ and $c_1=5.0$. }
\label{Fig2}
\end{figure}

As before, we consider a double-quench process and introduce three Hamiltonians with the corresponding parameters respectively labelled by $m_i$ and $c_i$ ($i=0,1,2$).
If all $c_i$s are set to $1.0$, the result has no physical difference from the previous SSH model. Here we consider the more general situation.
The existence of a critical momentum $k_\text{c}$ for the metamorphic DQPT requires
\begin{eqnarray}\label{KCe}
& &\hat{n}_{1k}\cdot\hat{n}_{2k} \notag \\&=&\frac{4M_{1}M_{2}}{\Delta_{1k}\Delta_{2k}}\left[\sin^{2} k
+(m_{1}-c_{1}\cos k)(m_{2}-c_{2}\cos k)\right],\notag\\
\end{eqnarray}
Let $y=\cos k_\text{c}$ and solve Eq. (\ref{KCe}), we get
\begin{align}\label{kc1}
y=\frac{(m_{1}c_{2}+m_{2}c_{1})\pm \Delta^2}{2(c_{1}c_{2}-1)}.
\end{align}
where $\Delta=\sqrt{(m_{1}c_{2}-m_{2}c_{1})^2-4c_{1}c_{2}+4+4m_{1}m_{2}}\geqslant 0$, as required by Vieta's theorem. The constraint $-1 \leqslant y \leqslant 1$ imposes more stringent conditions of the parameters $m_1, m_2, c_1$, and $c_2$.
Furthermore, the requirement $\hat{n}_{0k_\text{c}}\parallel\hat{n}_{2k_\text{c}}$ is satisfied if $m_2-m_0=(c_2-c_0)\cos k_{\text{c}}$. When all those conditions are satisfied, a metamorphic DQPT can occur in the double-quench process if the duration between the two quenches is
\begin{align}\label{kc2}
\tau^*=\frac{n\pi+\frac{\pi}{2}}{M_1\sqrt{(c_1\cos k_c-m_1)^2+\sin^2 k_c}}.
\end{align}

\begin{figure}[th]
\centering
\includegraphics[width=3.2in,clip]{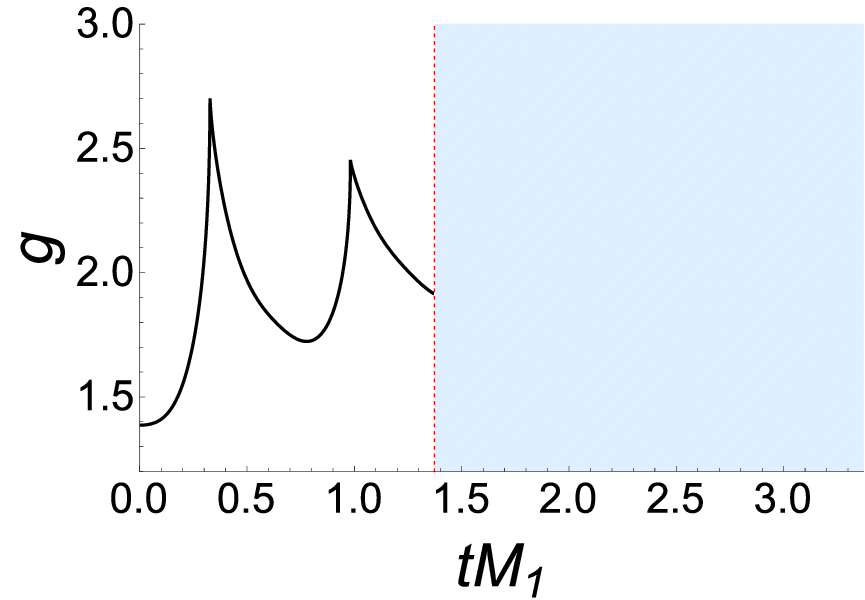}
\includegraphics[width=3.2in,clip]{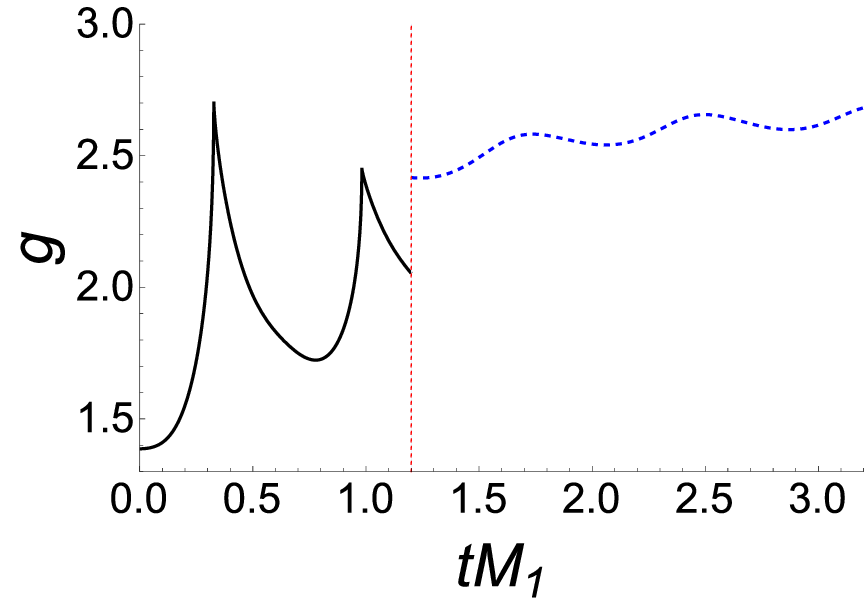}
\caption{(Color online ) $g(t)$ as a function of $t$ for the Kitaev chain with $M_1=M_2=1.0$, $m_1=0.2$, $c_{1}=5.0$, and $m_2=c_2=2.0$ at $T=5.0M_1$. (Top) Time duration $\tau=\tau^*_2=1.373/M_1$ (the red solid line) with a metamorphic
DQPT occurring at $t=\tau^*_2$. $g(t)$ remains singular after the second quench. (Bottom) Time duration $\tau=1.2/M_1$ (indicated by the vertical line) without a metamorphic DQPT. When $\tau\neq\tau^*_{1,2}$, $g(t)$ remains regular after the second quench. The kinks on both panels in the range $0<t<\tau$ indicate the ordinary DQPTs.}
\label{Fig3}
\end{figure}

Altogether, there are four parameters of the Kitaev chain that control the existence of metamorphic DQPTs. To simplify the discussion, we fix Hamiltonian after the first quench and search for suitable parameters of the Hamiltonian after the second quench. For example, if we set $m_{1}=1.0$ and $c_{1}=1.0$, the roots of Eq. (\ref{kc1}) are $y_1=\frac{m_2+1}{c_2-1}$ and $y_{2}=1$.
Thus, there always exists a critical momentum $k_\text{c}=2n\pi$ for all possible choices of $m_2$ and $c_2$ at any temperature. Next, we try another choice: $m_1=0.2$ and $c_{1}=5.0$. For this set of parameters, the existence of $k_\text{c}$ depends on the choices of $m_2$ and $c_2$ in Eq. (\ref{kc1}).
In Figure \ref{Fig2}, we show a $c_2-m_2$ phase diagram with the shaded region indicating where metamorphic DQPTs may occur. Different from the SSH model, there exist two possible critical momenta for the Kitaev chain in this case: \begin{equation}k_{\text{c}1,2}=\arccos\frac{\left(\frac{c_2}{5}+5m_2\right)\pm \Delta^2}{2(5c_2-1)}.\end{equation}
As a consequence, there are two possible choices of $\tau^*$, labeled as $\tau^*_{1,2}$, at which metamorphic DQPTs may happen according to Eq. (\ref{kc2}). We remark that the topological transition point of the Kitaev chain is at $|m|=c$~\cite{Kitaev}, so the double quenches for the metamorphic DQPTs are not tied to the topological transition point, in contrast to the metamorphic DQPTs of the SSH model.

To understand these results more clearly, we choose an example of the Kitaev chain with $M_1=M_2=1.0$, $m_1=0.2$, $c_{1}=5.0$, and $m_2=c_2=2.0$ at $T=5.0M_1$ and present our numerical results in Fig. \ref{Fig3}. In the top panel, $g(t)$ is plotted as a function of $t$, and the time duration between the two quenches is set to $\tau=\tau^*_2=1.373/M_1$, where $\tau^*_2$ is obtained from Eq. (\ref{kc2}) by setting $n=0$ and using $k_{\text{c}2}=\arccos\frac{\left(\frac{c_2}{5}+5m_2\right)-\Delta^2}{2(5c_2-1)}$. In this case, a metamorphic DQPT occurs at $t=\tau^*_2$. When $t<\tau^*_2$, there are also two ordinary DQPTs due to the first quench, which are actually related to the other critical momentum $k_{\text{c}1}=\arccos\frac{\left(\frac{c_2}{5}+5m_2\right)+\Delta^2}{2(5c_2-1)}$ via $n=0, 1$ in Eq. (\ref{t1n}), respectively. When $t>\tau^*_2$, the behavior of $g(t)$ is concealed by the metamorphic DQPT at $\tau^*_2$ as $g(t)$ becomes singular after the second quench.
In the bottom panel, we choose a different duration $\tau=1.2/M_1$, which does not match the condition of metamorphic DQPTs. Therefore, the behavior of the rate function for $t>\tau$ is regular and visible. As discussed before, no ordinary DQPT arises after the second quench in the latter case.

\begin{figure}[th]
\centering
\includegraphics[width=3.2in,clip]{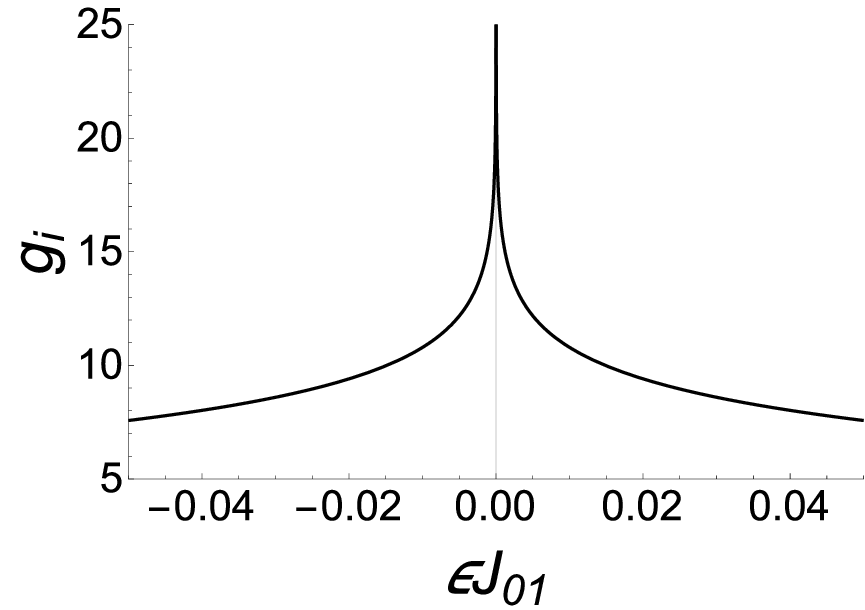}
\caption{The singular contribution $g_i$ to the rate function as a function of the time deviation $\epsilon$ from $\tau^*$ for the SSH model in the double-quench process depicted in Fig.~\ref{Fig1}. }
\label{Fig5}
\end{figure}

\section{Experimental implications}\label{sec:4}
Ref.~\cite{DQPTreview18} summarizes some pioneering experimental realizations of DQPTs from single-quench processes in trapped ions or ultracold atoms. Moreover, a direct measurement of the nonanalytic behavior of the rate function has been carried out in a simulator of interacting transverse-field Ising model in Ref. \cite{DQPTB41} and in topological nanomechanical systems in Ref.~\cite{Tian19}. There are more recent demonstrations of DQPTs and their implications in correlation functions~\cite{DQPTN17a,DQPTN17b}, spinor condensates~\cite{Yang19, Duan20}, photonic platforms~\cite{WangPRL19}, superconducting qubits~\cite{GuoApplied19}, NV centers in diamonds~\cite{ChenAPL20}, and nuclear magnetic resonance quantum simulators~\cite{Nie20}. In addition, observation of DQPTs through the dynamical vortices after a sudden quench close to a topological phase transition has been reported~\cite{DQPTB4}, which may be interpreted as the Fisher zeros of the Loschmidt amplitude~\cite{PhysRevLett.118.180601}. Since double-quench processes can be performed in similar fashions by including a second quench of the Hamiltonian, the metamorphic DQPT should be realizable in similar platforms that demonstrate the ordinary DQPTs. Therefore, the predictions of the metamorphic DQPT in this work should be experimentally verifiable. For example, the final state after a metamorphic DQPT will never return to the initial state, making the measurable rate function~\cite{DQPTreview18,DQPTB41,Tian19} singular after the second quench.

Theoretically, the rate function $g(t)$ will stay divergent if the second quench is applied exactly at $t=\tau^*$, as we have discussed before. In realistic situations, however, the time duration between the two quenches may not be exactly $\tau=\tau^*$ in experiments, where a small deviation $\epsilon$ from $\tau^*$ may arise. According to Eq. (\ref{gt}), the dominant contribution to $g(t)$ after the second quench in the case with a metamorphic DQPT is the singular term $\ln|\cos\omega_{1k_\text{c}}\tau^*|$. To estimate the influence of the deviation from $\tau^*$ in experiments, we plot $g_i=-\frac{2}{N}\ln|\cos\omega_{1k_\text{c}}(\tau^*+\epsilon)|$ vs. $\epsilon$ for the SSH model in Fig. \ref{Fig5}, following the double-quench process depicted in Fig~\ref{Fig1}. For a small deviation $\epsilon$, we found $g_i\sim -\ln|\sin\omega_{1k_\text{c}}\epsilon|\sim -\ln|\epsilon|$.  For a system with a metamorphic DQPT, the logarithmic divergence of the rate function as a function of the deviation of the duration between the two quenches thus serves as another signature of the metamorphic DQPT. A similar analysis of the Kitaev chain in the double-quench process shown in  Fig.~\ref{Fig3} exhibits a similar logarithmic divergence with the deviation $\epsilon$ from $\tau^*$ as well.
We also remark that the 1D two-band models analyzed here allows their energy spectra to be characterized by a periodic parameter (momentum $k$ for example). If more complicated interactions are introduced, the construction and analysis of the Loschmidt amplitude may depend on the details of the system, making it challenging to predict whether a metamorphic DQPT arises.

\section{Conclusion}\label{sec:con}
We have analyzed the dynamical behavior of generic 1D two-band systems going through double-quench processes at finite temperatures and presented a type of DQPT, named the metamorphic DQPT, where the final state continually has no overlap with the initial state. For the 1D two-band systems analyzed here, the metamorphic DQPT is not possible in single-quench processes. The general conditions for the existence of metamorphic DQPTs are derived. We discuss the implications of the metamorphic DQPT in two examples. In both the SSH model and Kitaev model, suitable choices of the quench Hamiltonians and the duration between the quenches can induce a metamorphic DQPT at the second quench, causing the Loschmidt amplitude to vanish and the rate function to be singular after the second quench. The general formalism applies to pure states at zero temperature as well as mixed states at finite temperatures. Our findings help provide more controls of dynamical evolution of quantum systems in future experiments, possibly including strongly  interacting systems~\cite{Peotta21,Brange22}.

\begin{acknowledgments}
H. G. was supported by the National Natural Science Foundation
of China (Grant No. 12074064). C. C. C. was supported by the National Science Foundation under Grant No. PHY-2011360.
\end{acknowledgments}

\bibliographystyle{apsrev}

\end{document}